\documentclass[letterpaper, twocolumn, 10pt]{article}
\usepackage{graphicx} 
\usepackage[right=1in, left=1in, top=1in, bottom=1in]{geometry}
\usepackage{amsmath}
\usepackage{amssymb}
\usepackage{amsfonts}
\usepackage{fdsymbol}
\usepackage{authblk}
\usepackage[colorlinks=true,citecolor=cyan,linkcolor=red,urlcolor=Green,bookmarks=true,bookmarksnumbered=true]{hyperref}
\usepackage{cite}
\usepackage[normalem]{ulem}
\usepackage[dvipsnames]{xcolor}

\begin{document}

\title{\bf\huge Low-Energy DNA Bubble Dynamics via the Quantum Coulomb Potential}
\author{Juan D. García-Muñoz$^1$, A. Contreras-Astorga$^2$ and L. M. Nieto$^3$}
\affil{\small $^1$Physics Department, Cinvestav, Av. Instituto Politécnico Nacional 2508, 07360 Mexico City, Mexico. \\
$^2$SECIHTI - Physics Department, Cinvestav, P.O. Box 14-740, 07000, Mexico City, Mexico. \\
$^3$ Departamento de F\'{\i}sica Te\'orica, At\'omica y Optica and Laboratory for Disruptive Interdisciplinary Science (LaDIS), Universidad de Valladolid, 47011 Valladolid, Spain.
}
\affil{\footnotesize E-mail: juan.garciam@cinvestav.mx, alonso.contreras@cinvestav.mx, luismiguel.nieto.calzada@uva.es}
\date{}

\twocolumn[

\maketitle

\vspace{-5\baselineskip}

\begin{abstract}
    {\bf Abstract:}  

We developed a low-energy model that can be used at any time to describe the dynamics of DNA bubbles at temperatures below the melting point.    
The Schrödinger equation associated with this problem is solved in imaginary time with a quantum Coulomb potential, and we obtain an approximate expression for its more general physical solution as a linear combination of the states whose energies are close to the lower bound energy.
We can then determine the probability density, the first-passage time density, and the correlation functions in terms of Bessel functions. Our findings are consistent with results obtained directly from the Fokker-Planck equation. Comparisons with the Gamma and Diffusion models are discussed.    \\
    
    \vspace{0.25\baselineskip}
    \begin{small}
    \end{small}
\end{abstract}

\vspace{2\baselineskip}
]

\section{Introduction}
DNA Denaturation consists of the breaking of base pairs in the DNA molecule through thermal activation, a process in which only a region of the molecule, known as the bubble, is generally broken.
In 1966, Poland and Scheraga constructed a model (PS) for the canonical partition function of DNA denaturation \cite{PS1966}, based on which there is now a large literature studying this phenomenon, with some interesting works appearing in Refs. \cite{WARTELL1985,HM2003,Bicout2004,Metzler2005,Tobias2007,FM2007a,FM2007b}. However, the PS model is not the only proposal to study denaturation, another alternative option being the Peyrard and Bishop model, which offers similar results \cite{Peyrand1989}.
In general, the dynamics of DNA bubbles are usually determined by a continuity equation 
$$
\frac{\partial P(x,t)}{\partial t} + \frac{\partial j(x,t)}{\partial x}=0,
$$ 
with the following expression for the probability density current, proposed from the polymer translocation model \cite{Chuang2001} 
$$
j(x,t) = -D\left(\frac{\partial P(x,t)}{\partial x} + \frac{P(x,t)}{k_BT}\frac{\partial\mathcal{F}}{\partial x}\right),
$$
where $D$ is the kinetic coefficient (setting the time scale, as $[D]=s^{-1}$), $x>0$ is the size of the bubble in the DNA molecule in terms of number of broken base pairs, $T$ is the temperature, $k_B$ is the Boltzmann constant, $P(x,t)$ is the probability density function of having a bubble of size $x$ at time $t$, and $\mathcal{F}$ is the free energy of the bubble, which is known to be given by $\mathcal{F}=k_BT(\gamma_0 + x\gamma + c\ln x)$ \cite{FM2007a}, where $\gamma(T) = \gamma_1(1-T/T_m)$, $\gamma_0\approx10(T_r/T)$ and $\gamma_1\approx4(T_r/T)$, values taken near the reference temperature $T_r\approx 37\ ^{\circ}$C and the melting temperature $T_m\approx 66\ ^{\circ}$C, while $c$ is a dimensionless constant having two values typically reported in literature, namely, $1.76$ and $2.12$ \cite{HM2003}.

\section{From the Fokker-Planck to the Schrödinger equation}

The bubble behavior in DNA can be described by the probability density function $P(x,x_0,t)$ which satisfies the Fokker-Planck equation (derived from the continuity equation) in the form
\begin{equation} \label{E2.1}
    \frac{\partial P}{\partial t} = 2\frac{\partial}{\partial x}\left(\frac{\mu}{x}-\epsilon\right)P+\frac{\partial^2 P}{\partial x^2},
\end{equation}
where $x_0$ is the initial bubble size, $\mu=c/2$ and $\epsilon=(\gamma_1/2)(T/T_m-1)$. It should be noted that time is rescaled according to $t\rightarrow Dt$ \cite{FM2007a,FM2007b,HM2003}. 

The Fokker-Planck equation in \eqref{E2.1} can be converted to a Schrödinger-type equation by transformation $P = e^{\epsilon x}x^{-\mu}\Psi(x,t)$, which leads to
\begin{equation} \label{E2.2}
    -\frac{\partial\Psi}{\partial t} = -\frac{\partial^2 \Psi}{\partial x^2} + \left(\frac{\mu(\mu+1)}{x^2}-2\frac{\mu\epsilon}{x}+\epsilon^2\right)\Psi.
\end{equation}
Note that this is an imaginary-time Schrödinger-type equation, with a Coulomb potential $2\mu\epsilon/x$, a centrifugal barrier $\mu(\mu+1)/x^2$, and an energy shift $\epsilon^2$. Its solution can be found in most Quantum Mechanics textbooks \cite{Cohen77,flugge94}. To solve this equation, we first assume that the time behavior of the solutions $\Psi$ is standard, {\it i.e.,} $\Psi(x,t) = e^{-Et}\psi(x)$, where $E$ is the corresponding energy eigenvalue of the solution $\psi(x)$ that satisfies the stationary Schrödinger equation
\begin{equation} \label{E2.3}
    -\frac{d^2 \psi}{d x^2} + \left(\frac{\mu(\mu+1)}{x^2}-2\frac{\mu\epsilon}{x}+\epsilon^2\right)\psi = E\psi.
\end{equation}
Due to their definitions, the parameter $\mu>0$, but $\epsilon\in\mathbb{R}$, so we have three possible cases.
i) If the parameter $\epsilon>0$, then the potential term in Eq.~\eqref{E2.3} admits bound states, but for this case the temperature $T>T_m$, and as was well studied in Refs. \cite{FM2007a,FM2007b}, the bubbles will continue to grow until the DNA is completely denatured, this being a process that is beyond the scope of this article.
ii) In the case of $\epsilon=0$, there is a critical behavior similar to an unbiased random walk.
iii) The most interesting case occurs when $\epsilon<0$, $T<T_m$, since bubbles can grow or contract until they finally close after a finite time. However, their behavior at any instant has not been adequately analyzed from the Schrödinger perspective just outlined, so we will first focus on finding a suitable solution $\psi(k,x)$ that will allow us to construct the probability density function $P(x,x_0,t)$ as follows:
\begin{equation} \label{E2.4}
    P(x,x_0,t) = Nx^{-\mu}e^{\epsilon x}e^{-\epsilon^2t}\int_{0}^{\infty}dk\ e^{-k^2t}C(k)\psi(k,x),
\end{equation}
with $k^2 = E-\epsilon^2$, $C(k)$ are the coefficients to be determined and $N$ the normalization constant. For an energy eigenvalue $E>\epsilon^2$, $\epsilon<0$ and $\mu>0$, the oscillatory solutions that converge for $x=0$ are given by
\begin{equation} \label{E2.5}
    \psi(k,x)=(i2kx)^{\mu+1}e^{ikx}F(a(k),2\mu+2,i2kx),
\end{equation}
where $a(k) = \mu+1-i\mu|\epsilon|/k$ and $F(a,b,z)$ is the confluent hypergeometric function \cite{Abramowitz1965}. Since the probability density function must satisfy 
$$
P(x,x_0,t=0) = \delta(x-x_0), 
$$
a natural choice of the unknown coefficients in \eqref{E2.4} is $C(k) = x_0^{-\mu}e^{\epsilon x_0}\bar{\psi}(k,x_0)$, where $\bar{z}$ denotes complex conjugation \cite{FM2007b}. Therefore, $P(x,x_0,t)$ turns out to be
\begin{equation} \label{E2.6}
    \begin{aligned}
        P(x,x_0,t)=&N(xx_0)^{-\mu}e^{\epsilon(x+x_0)}e^{-\epsilon^2t}\\
        &\times\int_{0}^{\infty}dk\, e^{-k^2t}\ \overline{\psi}(k,x_0)\, \psi(k,x). 
    \end{aligned} 
\end{equation}
Since the solutions $\Psi(x,t)$ decay exponentially with time, the functions $\psi(k,x)$ that contribute the most to the integral over $k$ are expected to be those whose energy eigenvalues are close to the lower bound $\epsilon^2$, although it is not easy to propose a valid approximation that allows us to perform the integral analytically in equation ~\eqref{E2.6}. Now, if we make the change of variable $z^2=k^2t$, the hypergeometric parameter $a(z)$ can be written as $a(z) = \mu+1-i\mu|\epsilon|\sqrt{t}/z$ and if we consider that $t/z^2\ll 1/\mu^2\epsilon^2$, then it reduces to the approximate form $a\approx\mu+1$. Note that this is a valid low-energy approximation for $E\ll (1+\mu^2)\epsilon^2$. Consequently, the probability density function turns out to be 
\begin{eqnarray} 
       P(x,x_0,t) \!\!\!&\!\approx\!&\!\!\! N4^{2\mu+1}\Gamma^{2}\left(\mu+\frac{3}{2}\right)(xx_0)^{-\mu+1/2}e^{\epsilon(x+x_0)}\nonumber  \\
       \!\!\!&\!\!&\!\!\! \times\frac{e^{-\frac{(x^2+x_0^2)}{16t}}e^{-\epsilon^2t}}{t}\ I_{\mu+1/2}\left(\frac{xx_0}{8t}\right), 
       \label{E2.7}
\end{eqnarray}
where $\Gamma(z)$ is the gamma function and $I_\nu(z)$ is the modified Bessel function of the first kind. It is worth mentioning that the above form of $P(x,x_0,t)$ must be normalized, i.e., $\int_{0}^{\infty}dx\ P(x,x_0,t=0)=1$. Using the asymptotic form of the modified Bessel function for large arguments, $I_\nu(z)\sim e^{z}/\sqrt{2\pi z}$ \cite{Abramowitz1965}, we obtain an asymptotic form for the probability density function at short times, explicitly
\begin{eqnarray}
        P(x,x_0,t)\!\!\!&\!\sim \!&\!\!\! N4^{2\mu+3}\Gamma^{2}\left(\mu+\frac{3}{2}\right)\nonumber  \\
        \!\!\!&\!\!&\!\!\!\times (xx_0)^{-\mu}e^{\epsilon(x+x_0)}\frac{e^{-\frac{(x-x_0)^2}{16t}}}{8\sqrt{\pi t}}.
         \label{E2.8}
\end{eqnarray}
\begin{figure*}[t]
    \centering
    \includegraphics[scale=0.8]{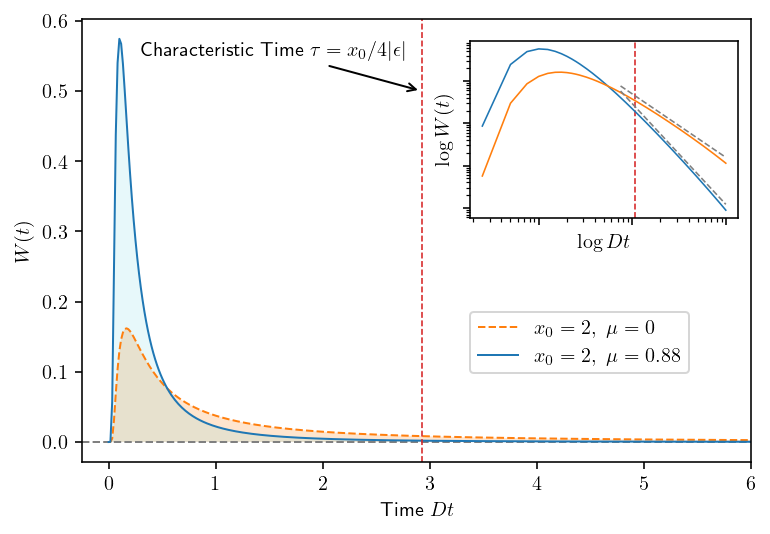}
    \caption{First-passage time density functions for the Bessel model in Eq. \eqref{E2.11} at $37\, ^{\circ}$C and different values of the parameter $\mu$. The inset graph shows a log-log plot of these functions and two straight gray dashed lines with slopes $-1.5$ and $-2.38$, respectively.}
    \label{F1}
    \end{figure*}
It is possible to evaluate the above equation at time $t=0$ and since $\lambda e^{-\lambda^2x}\rightarrow 2\sqrt{\pi}\delta(x)$ when $\lambda\rightarrow \infty$, 
the integral can be evaluated over the entire range of positions, finding the value of the normalization constant:
\begin{equation} \label{E2.9}
    N = \frac{x_0^{2\mu}e^{-2\epsilon x_0}}{\left[ \Gamma\left(\mu+\frac{3}{2}\right)\right]^{2} 4^{2\mu+3}}.
\end{equation}
We can determine an asymptotic form at large times for the probability density function. Taking the asymptotic form of the modified Bessel function for small arguments $I_\nu(z)\sim (z/2)^\nu/\Gamma(\nu+1)$, we have that
\begin{equation}\label{E2.10}
    P(x,x_0,t)\sim N\Gamma\left(\mu+\frac{3}{2}\right)xx_0e^{\epsilon(x+x_0)}\frac{e^{-\epsilon^2t}}{t^{\mu+3/2}}.
\end{equation}
It is worth mentioning that this large-time form of the probability density function is consistent with the results obtained in Refs. \cite{FM2007a,FM2007b}, where a large-time expression, called the Gamma model, is also obtained.
However, a better approximation of $P(x,x_0,t)$ for any time is the expression in equation ~\eqref{E2.7}, which we will call the \emph{Bessel model}. Furthermore, it is consistent with the expression for the critical case ($\epsilon=0$) given in the references mentioned above, which can be calculated without approximations.

\begin{figure*}[t]
    \centering
    \includegraphics[scale=0.8]{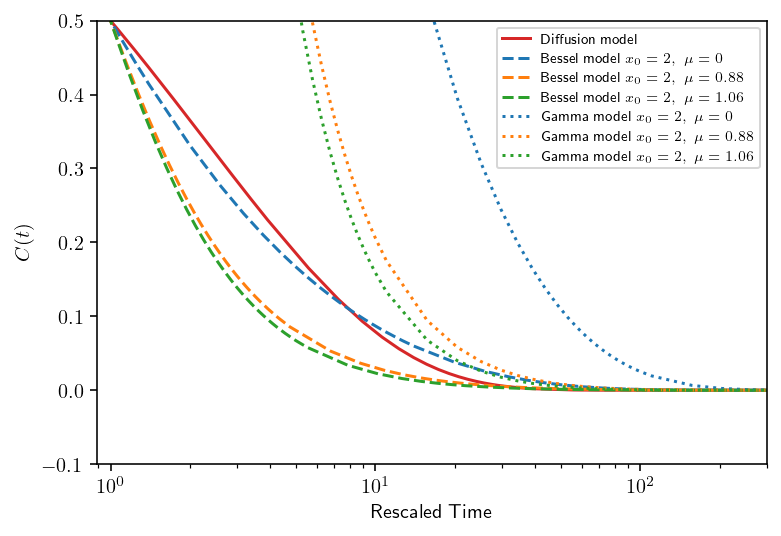}
    \caption{Comparative graph between the correlations of the Bessel model and the Gamma model, in contrast to the Diffusion model of Ref. \cite{AltanBonnet2003} at $37\, ^{\circ}$C. 
    The Bessel model shows a similar behavior to the Diffusion model with the parameter values set $x_0=2,\ \mu=0$. The correlations are centered at time $t_{1/2}$, so that $C(t_{1/2})=1/2$.}
    \label{F2}
\end{figure*}

On the other hand, in the analysis of bubble dynamics, the `first-passage time density' is important, defined as $W(t)=\int_0^{\infty}dx\ (\partial P/\partial t)$. However, since $P(x,x_0,t)$ satisfies the Fokker-Planck equation \eqref{E2.1}, and from equation \eqref{E2.7}, it can be seen that the probability density function and its derivative vanish when $x\rightarrow\infty$, so the first-passage time density can be written as $W(t) = (\partial P/\partial x)|_{x=0}+2(\mu/x-\epsilon)P|_{x=0}$, and takes the following form:
\begin{equation}\label{E2.11}
    W(t) = N(2\mu+1)\Gamma\left(\mu+\frac{3}{2}\right)x_0e^{\epsilon x_0}\frac{e^{-\frac{x_0^2}{16t}}e^{-\epsilon^2t}}{t^{\mu+3/2}}.
\end{equation}
It should be noted that this expression for the first-passage time density agrees with the one obtained in Ref. \cite{HM2003}, where the result is determined directly from the Fokker-Planck equation, both expressions decaying exponentially with exponent $(\alpha +\beta t)^2/4t$, with $\alpha,\beta$ being constant factors in terms of the initial bubble size $x_0$ and the parameter $\gamma(T)$.
Figure \ref{F1} shows the first-passage time density functions $W(t)$ in Eq.~\eqref{E2.11} for a pair of parameter sets: 
the set $\{x_0=2,\ \mu=0\ (c=0)\}$ approximates the Diffusion model \cite{AltanBonnet2003} (see Figure~\ref{F2}) and the set $\{x_0=2,\ \mu=0.88\ (c=1.76)\}$ corresponds to the typical value of $c$. The inset 
graph shows log-log plots of these density functions, where the gray dashed lines (corresponding to power function behavior) contrast with exponential behavior that can be observed.
Furthermore, we can know the so-called characteristic first-passage time for bubble closure, which, in the case of the Bessel model for $W(t)$, turns out to be $\tau = x_0/4|\epsilon|$.

To establish a connection with the experimental data, an important quantity is the correlation function, defined as $C(t)=1-\int_{0}^{t}W(s)ds = \int_{t}^{\infty}W(s)ds$. The Bessel model used here gives the following approximate expression for the correlation function:
\begin{eqnarray} 
C(t) \!\!\!&\!=\! & \!\!\! N\, 4^{\mu+1}\,  (2\mu+1)\, \Gamma(\mu+3/2)\ e^{\epsilon x_0}\, \frac{|\epsilon|^{\mu+1/2}}{x_0^{\mu-1/2}}\nonumber   \\
     \!\!\!&\!\!& \!\!\!   \times\,  K_{\mu+1/2}\left(\epsilon^2t,\frac{x_0|\epsilon|}{2}\right),  \label{E2.12}
\end{eqnarray}
where $K_\nu(a,z) =\frac{1}{2} (z/2)^\nu \int_{a}^{\infty}e^{-\left(\frac{z^2}{4s}+s\right)}\frac{ds}{s^{\nu+1}}$ is the incomplete modified Bessel function of the second kind~\cite{NIST}. 

In Fig. \ref{F2} we show a comparison between the correlations of the Bessel model and the Gamma model, contrasting with the Diffusion model, which is considered an adequate fit to the experimental data presented in Ref. \cite{AltanBonnet2003}.
In that figure, three sets of parameter values are plotted: for the first one $\{x_0=2,\, \mu=0\ (c=0)\}$, the Bessel model is close to the Diffusion model; for the second one $\{x_0=2,\, \mu=0.88\ (c=1.76)\}$, our model shows a clear difference respect to the Diffusion model; and for the third one $\{x_0=2,\, \mu=1.06\ (c=2.12)\}$, the correlation function behaves similar to the previous case.
As we can see, when the parameter $\mu$ is close to zero, the Bessel model and the Diffusion model behave similarly, implying that the parameter $c\approx 0$, which is not surprising since the Diffusion model does not consider the loop entropy term in the free energy of a bubble \cite{WARTELL1985,HM2003,Metzler2005}. Moreover, since $I_{1/2}(z)=\sqrt{2/(\pi z)}\sinh{z}$ \cite{NIST}, the Bessel model has a well-defined limit when $\mu\rightarrow 0$ for the probability density function in Eq. \eqref{E2.7} as well as for the first-passage time density in Eq. \eqref{E2.11} and for the correlation function in Eq. \eqref{E2.12}, which turns out to be written in terms of complementary error functions erfc$(z) = 2/\sqrt{\pi}\int_z^\infty e^{-s^2}ds$. Since the Bessel model is a low-energy approximation, in this case, $E\rightarrow\epsilon^2$. We must mention that, when $\mu=0$, the Schrödinger-type equation \eqref{E2.3} reduces to the free particle case and its solutions simplify to $\psi(x)\propto e^{i2kx}\sin kx$. Although this case can be easily addressed, it is not of interest, since the literature reports positive values of $c$, for example, in reference \cite{WARTELL1985} it is reported that $1.5\le c\le 2$, and in reference \cite{Kafri2000} it is indicated that $c=2.12$.

\section{Conclusions}

A suitable expression for the probability density function describing the dynamics of DNA bubbles in terms of Bessel functions was determined. This expression is a low-energy approximation valid for any time, in the case of temperatures below the melting temperature $T_m$, while, in contrast, the results of the Gamma model in \cite{FM2007a,FM2007b} only provide an expression for long times.
The first-passage time density function \eqref{E2.11} obtained from the Bessel model replicates the exponential decay observed for the results derived directly from the Fokker-Planck equation \eqref{E2.1} and leads to a characteristic first-passage time $\tau=x_0/4|\epsilon|$. The correlation function of the Bessel model \eqref{E2.12} exhibits a behavior similar to that of the diffusion model when the parameter $\mu\approx0$. However, our model can be used to fit experimental data, such as those shown in \cite{AltanBonnet2003}, and determine a value for the coefficient $c$ of the loop entropy term in the bubble free energy. Finally, it is worth mentioning that our findings can be generalized to other profiles of the $\gamma(T)$ function, such as those proposed in \cite{Wu2009}, which constitutes a very interesting work for the future, on which we are already working.

\section*{Acknowledgments}
JDGM acknowledges support from SECIHTI for the postdoctoral fellowship with CVU number 712251 and the facilities provided by the Universidad de Va\-lla\-dolid for the completion of this work. The research of LMN and JDGM was supported by the Q-CAYLE project, funded by the European Union-Next Generation UE/MICIU/Plan de Recuperacion, Transformacion y Resiliencia/Junta de Castilla y Leon (PRTRC17.11), by the project PID2023-148409NB-I00, funded by MICIU/AEI/10.13039/501100011033, and also by the Department of Education of the Junta de Castilla y Leon and FEDER Funds (Reference: CLU-2023-1-05). ACA and JDGM acknowledge Se\-cretar\'ia de Ciencia, Humanidades, Tecnolog\'ia e Innovaci\'on (SECIHTI - M\'exico) support under the grant FORDECYT-PRONACES/61533/2020.

 \bibliography{References}
 \bibliographystyle{rmf-style}
\end{document}